# Optimized Multi-Party Quantum Clock Synchronization

Radel Ben-Av and Iaakov Exman

Software Engineering Department
Jerusalem College of Engineering
POB 3566, Jerusalem, 91035, Israel
rbenav@gmail.com, iaakov@jce.ac.il

**Abstract.** A multi-party protocol for distributed quantum clock synchronization has been claimed to provide universal limits on the clock accuracy, viz. that accuracy monotonically decreases with the number *n* of party members. But, this is only true for synchronization when one limits oneself to W-states. This work shows that usage of *Zen*-states, a generalization of W-states, results in improved accuracy, having a maximum when $\lfloor n/2 \rfloor$ of its members have their qubits with a |1> value.

## 1 Introduction

Entanglement as a physical resource has been extensively investigated for a variety of applications in distributed systems – for instance, QKD (quantum key distribution) and QCS (quantum clock synchronization). QCS Protocols were published referring to synchronization of a pair of clocks and later to the multi-party case.

In the introduction we describe basic quantum clock synchronization ideas and the multi-party protocol proposed by Krco and Paul [13] based upon W-states. In the remainder of the paper, we introduce *Zen*-states as a generalization of W-states (in section 2), use Zen-state properties to show that one can optimize multi-party quantum clock synchronization beyond the Krco and Paul protocol (in section 3), and conclude with a discussion (in section 4).

### 1.1 Essentials of Quantum Clock Synchronization

Quantum clock synchronization is based upon preparation of entangled states, to be later used for synchronization. Once entangled states are prepared, measurements may be performed triggering a well-defined time evolution between these states, enabling synchronization up to a given precision.

Josza et al. [12] proposed a two-party quantum clock synchronization protocol requiring a shared entangled state and classical messages between the two parties, in which the classical messages do not carry timing information.

### 1.2 Multi-Party Clock Synchronization Protocols

Krco and Paul [13] extended the Josza et al. two-party protocol to a multi-party synchronization protocol. Its purpose is to synchronize *n* spatially separated clocks, any one of which can be later taken as the standard clock. The protocol starts from an initial W-state (see e.g., Dur et al. [7], although they are not called as such in Krco and Paul's paper). W-states are entangled states which have *n* terms in the following form:

$$|W(N)> = (|100...0> + |010...0> + |001..0> + \cdots |000...1>) \qquad (1)$$

Each of these terms contains a single qubit in the |1> state. The initial W-state |W> is an energy eigenstate, since |0> and |1> are themselves assumed to be energy eigenstates with different energies. This ensures that |W> is invariant until measurements are made.

At standard time $t_A=0$, Alice – which has the standard clock – measures the qubit in her possession in the measurement (|+>,|->) basis. She then publishes classically the measurement results. Bob – a

generic name for the holder of a clock to be synchronized – also measures his qubits in the measurement basis, at time $t_B$, which is skewed by Δt from the standard time.

Following the application of the time evolution operator, for sets measured by Alice as |+>, Bob gets the probabilities *P* of its two possible outcomes:

$$P(|\pm>) = \frac{1}{2} \pm \left(\frac{\cos(\omega\Delta t)}{n}\right) \qquad (2)$$

Assuming that $|\omega\Delta t|<2\pi$, Bob's measurements allow him to estimate Δt and adjust his clock.

Krco and Paul in the analysis of their result state that the accuracy of determination of Δt decreases with *n*, since – following equation (2) above – the amplitude of the probability variation decreases with *n*. They attribute it to the decrease in entanglement with *n* of the initial state – in equation (1). Furthermore, while they state that it is worthwhile to look for a different initial state other than (1), they suggest that their limits are universal.

In the following section, we show that indeed a different initial state changes the view that the accuracy of determination of Δt decreases with *n*.

## 2  *Zen*-State: A Generalization of W-states

We start by defining the *Zen*-state[1] notation. *Zen*-state is shorthand for a fully symmetric entangled state with N qubits. It is fully symmetric under the operation of particle exchange. It is a generalization of W-states, as seen below.

### 2.1    Preliminaries: *Zen*-State Notation

A *Zen*-state is denoted by $|Z_k(N)>$ where N is the total number of qubits (particles) and k is the number of qubits in the |1> state in each term. It generalizes W-states for which there is the restriction that k=1.

The state |000…0> with N particles is $|Z_0(N)>$. It is the null-state of the *Zen*-state structure, as it has no entanglement (it is dual to $|Z_N(N)>$).

The first actually entangled example of a *Zen*-state is given by the (non-normalized) *Zen*-state $|Z_1(N)>$ of N particles with one particle in the |1> state – identical to a W-state – as follows:

$$|Z_1(N)> = (|100...0> + |010...0> + |001..0> + \cdots + |000...1>) \qquad (3)$$

The normalized Zen state[2] can be derived from the non-normalized one. For k=1:

$$|\tilde{Z}_1(N)> = \frac{1}{\sqrt{N}}|Z_1(N)> \qquad (4)$$

### 2.2    *Zen*-State Properties Relevant to Synchronization

Here we assume that for every single qubit the |1> state is in a different energy than the |0> state. Before setting up the clock we are interested in time invariant states, i.e. stationary, up to at most an overall time dependent phase. We naturally focus on states with well-defined energy (linear combination of states with definite numbers of |0>'s and |1>'s), since these are eigenfunctions of the Hamiltonian, i.e. they may be composed only of degenerate eigenvectors of the Hamiltonian. We only discuss states that are symmetric w.r.t. particle exchange as this is a natural property of our systems.

One can easily verify that for |0> and |1> states differing in energy – of relevance for time synchronization – the *Zen*-state $|Z_k(N)>$ is stationary for any k. Indeed there is a global time dependent phase, however since the quantum states are rays in the Hilbert space the overall total phase is irrelevant.

In what follows we shall consider *Zen*-states for any values of k.

---

[1] *Zen* is an abbreviation of *Z-en*tanglement, Z standing for time (Zeit in German). Zen states were also coined symmetric Dicke states [8].
[2] The tilde ~ above the Z letter, say $\tilde{Z}$, means that it is normalized.

# 3 Quantum Clock Synchronization Optimization

## 3.1 Optimization Idea

We follow the approach of Krco and Paul to multi-party synchronization. We relax the constraint that the number of |1> valued qubits per member of the initial state is exactly k=1. Thus our normalized initial state has *k* qubits with value |1> per member and (*n-k*) qubits with value |0>:

$$|\tilde{Z}_k(N)> = \left(\sqrt{\frac{n!}{(n-k)!k!}}\right)^{-1} * (|111...000> + |11...01...000> + \cdots + |000...111>) \quad (5)$$

The idea is very simple. Once the k=1 constraint is relaxed, one has an additional degree of freedom for optimization, viz. the variability of k, which can be optimized. Moreover, one can guess that since our systems display duality between |0> and |1> states, the optimal value of k is k=$\lfloor n/2 \rfloor$, as it will be shown in the next sub-sections.

## 3.2 Density Matrix Calculation for $Z_k(N)$ States

The density matrix calculations for $Z_k(N)$ states is outlined as follows:

1. **Density Matrix in Computational Basis** – obtain the partial density matrix in the computational basis for the standard clock A and a generic clock to be synchronized B. This is denoted $\rho^c_{AB}$ (the superscript *c* stands for computational, i.e. for qubit states |1> and |0>). This matrix is shown in the next equation:

$$\rho^c_{AB} = \frac{1}{n(n-1)} \begin{pmatrix} (n-k)(n-k-1) & 0 & 0 & 0 \\ 0 & k(n-k) & k(n-k) & 0 \\ 0 & k(n-k) & k(n-k) & 0 \\ 0 & 0 & 0 & k(k-1) \end{pmatrix} \quad (6)$$

2. **Density Matrix in Measurement Basis** – transform the previous partial density matrix into $\rho^m_{AB}$ (the superscript *m* stands for measurement states |+> and |->).
3. **Bob's Density Matrix** – obtain the density matrix of Bob $\rho^m_B$, corresponding to Alices's |+> states.
4. **Bob States' Probabilities** – using $\rho^m_B$ one finds that the probabilities of Bob states $P(\pm)$ are:

$$P(\pm) = \frac{1}{2} \pm \rho^c_{AB}(01,01) * \cos(\omega \Delta t) \quad (7)$$

hence

$$P(\pm) = \frac{1}{2} \pm \frac{k(n-k)}{n(n-1)} * \cos(\omega \Delta t) \quad (8)$$

## 3.3 Optimization Detailed

In order to improve the clock adjustment accuracy one chooses an optimal k for a given N as follows.
We denote by $A_0$ the amplitude of the time probability fluctuation. Hence:

$$A_0(k,n) = \frac{k*(n-k)}{n*(n-1)} \quad (9)$$

For any given *n* we wish to choose *k* such that $A_0$ is maximized.

One can easily see that $k_{opt}$ (optimal $k$) is

$$k_{opt} = \lfloor n/2 \rfloor \qquad (10)$$

for which

$$A_0(k_{opt}, n) = \frac{\lfloor n/2 \rfloor \lceil n/2 \rceil}{n*(n-1)} \qquad (11)$$

For n≥4 our result is a clear improvement over the original W-state (k=1), for which $A_0(1,n) = 1/n$.

### 3.4  Benefits and Limitations

The entanglement dependent coefficient $A_0(k,n)$ in eq. (9) is a symmetric function of k, with a maximal value in k=$\lfloor n/2 \rfloor$. The symmetry of $A_0$ is an expression of the duality between |0> and |1>.

In addition, if for some reason one cannot use the optimal k, one can still improve accuracy to a certain desirable extent by using an intermediate value of k.

It is noteworthy that for $k_{opt}$: $A_0 \xrightarrow[n\to\infty]{} \frac{1}{4}$, i.e. for the optimal choice of k the accuracy of the clock synchronization does not suffer a significant reduction as a function of N.

## 4  Discussion

There is an extensive literature on QCS (quantum clock synchronization). The following references are a representative sample of this literature.

### 4.1  Related Work

Josza et al. [12] is a basic reference for the synchronization of two spatially separated parties based upon shared prior quantum entanglement and classical communications. The accuracy of the protocol is independent of the two parties' knowledge of their relative locations or of the intervening medium properties.

Chuang [5] describes a quantum TQH (Ticking Qubit Handshake) protocol allowing two spatially separated clocks to be synchronized independently of the uncertainties in message transport time between them. This protocol requires O(*n*) quantum messages to obtain the *n* digits of the time deviation Δ between the clocks.

Optimization and limiting issues have been dealt with in the quantum clock literature, in particular with respect to QCS. Buzek et al. [4] have shown that as the dimension of the clock's Hilbert space grows to infinity, the time resolution bound given by the energy eigenvalues difference and the Holevo bound on classical information encoded by quantum means are satisfied simultaneously. Preskill [14] considers entanglement distillation and quantum error-correcting codes as ways to improve the robustness of QCS protocols. Giovannetti et al. [9] propose the combination of entanglement and squeezing of light pulses to enhance the accuracy of clock synchronization relative to classical protocols with light of the same frequency and power. Yurtsever and Dowling [16] present a relativistic analysis of the basic QCS protocol [12] and conclude that some method of entanglement distribution is needed to overcome entanglement purification issues. Harrelson and Kerenidis [10] dealt with similar issues. Burt et al. [3] in a reply to [12] state that it is essentially a kind of Eddington slow clock transfer synchronization protocol, while discussing its limitations, Boixo et al. [2] discuss decoherence in the context of quantum versions of the Eddington protocol, and ways to achieve the Heisenberg clock-synchronization limit. deBurgh and Bartlett [6] propose a method to achieve a better accuracy than the standard quantum limit without entanglement.

Janzing and Beth [11] refer to the synchronization of bipartite quantum clock systems by means of classical one-way communication and their thermodynamic implications.

Multi-party QCS protocols were first considered by Krco and Paul [13] based upon W-states, as already referred to in sub-section 1.2.

Experimental work has also been done on QCS implementation. Valencia et al. [15] report on a distant clock synchronization experiment (picosecond resolution at 3 kilometer distance) based upon entangled photon pairs. Bahder et al. [1] describe two party synchronization, based on second-order quantum interference between entangled photons generated by parametric down-conversion.

### 4.2    Open Issues

Future investigation may consider the issue of choosing optimal *k* in the presence of noise models.

It should be clear that the improved accuracy obtained does not solve the relative phases problem mentioned, e.g. in the Krco and Paul paper.

### 4.3    Main Contribution

The main contribution of this paper is to show that for multi-party synchronization, suitable choice of the initial entangled state – viz. Zen-states $Z_k(n)$, a generalization of W-states – improves the accuracy of quantum clock synchronization, over straightforward use of W-states. The improvement increases as the number of participants grows such that the accuracy of the time measurement does not depend on N for large enough number of participants.